\def\degree{\ifmmode^\circ\else${}^\circ$\fi}
\def\micron{\ifmmode\,\mu\else$\,\mu$\fi}

\documentclass[aps,prl,twocolumn,groupedaddress,amsmath,amssymb,showpacs]{revtex4}
\newcommand{\bsim}{\mbox{\raisebox{-0.1cm}{$\;
\stackrel{\textstyle>}{\sim}\;$}}}
\newcommand{\lsim}{\mbox{\raisebox{-0.1cm}{$\;
\stackrel{\textstyle<}{\sim}\;$}}}

\usepackage{graphicx}% Include figure files
\usepackage{dcolumn}% Align table columns on decimal point
\usepackage{bm}% bold math
\usepackage{psfig}% bold math

\begin{document}

%\keywords{}

%\maketitle


%\section{\label{Intro}Introduction}
\noindent{\textbf{ Comment on ``
Electron-phonon Vertex in the Two-dimensional One-band Hubbard Model''}}
\vskip\baselineskip

In the past months a variety
of experiments have pointed out
an important role of the electron-phonon (el-ph) interaction in many physical
properties of the
cuprates \cite{lanzara,keller,dastuto}.
These recent findings have triggered
a renewed interest for a theoretical understanding of the electron-phonon
properties in strongly correlated systems.

One of the most remarkable effects of the strong electronic correlation
on the el-ph properties is a to favor
forward (small ${\bf q}$) scattering
in the electron-phonon vertex, ${\bf q}$ being the exchanged
phonon momentum.
This feature was investigated in the past
by means of analytical techniques based on slave-bosons
or Hubbard $X$-operators \cite{grilli,zeyher,kulic}.
The assumption of forward scattering predominance within an el-ph
framework was shown to explain in a natural way
several anomalous properties of cuprates as the difference between transport
and superconducting el-ph coupling constants \cite{zeyher},
the linear temperature behaviour of the resistivity \cite{cp},
the $d$-wave symmetry or the superconducting gap \cite{kulic,pgp}.
Small ${\bf q}$ scattering selection was shown moreover to be
responsible in a natural way for high-$T_c$ superconductivity
within the context of the nonadiabatic superconductivity \cite{gpsprl}.

So far, this important feature was analyzed only by means of
analytical approaches in the $U = \infty$ limit
and a definitive confirm of it based
on numerical methods was lacking. With these motivations
in a recent paper Huang {\em et al} have address this issue
in the twodimensional Hubbard model with generic $U$
by using Quantum Monte Carlo (QMC) techniques
on a $8 \times 8$ cluster \cite{huang}.
Their results provide a good agreement with the previous analytical
studies and represent an important contribution to assess the relevance
of el-ph interaction in correlated system.

Moved by similar motivations, we have recently investigated
the onset of forward scattering predominance in the el-ph vertex
by using slave-boson techniques based
on four auxiliary fields \cite{cerruti}. While the previous
analytical studies
were limited to the $U = \infty$ limit \cite{grilli,zeyher,kulic},
we were able in this way
to evaluate the small momenta
selection in the whole phase diagram of parameters $U$-$n$,
where $n$ is the electron filling, in direct comparison with
Huang {\em et al.} \cite{huang}.
Our findings reproduce in more than a qualitative way the results
of Ref. \cite{huang}, and they will be object of a future
publication which is in preparation at the moment \cite{ccp}.

Besides the direct numerical confirmation of small momenta predominance
in correlated systems, Huang {\em et al.} report in their paper
an increasing of the el-ph vertex as function of $U$
in the large $U$ regime for small phonon momenta and a relative saturation
for large ${\bf q}$ \cite{huang}.
The origin of this puzzling feature has been not well
understood so far and a possible connection with charge excitations
driven by the exchange term $J \propto t^2/U$ in the large $U$ limit
was mentioned \cite{huang}.

In this Comment we would like to make clearer the issue.
In particular  our analytical calculations suggest that
the increase of the el-ph vertex function
at small ${\bf q}$ as function of $U$ is a peculiar feature of
the high temperature regime and it could be associated to
the tendency towards a phase-separation instability.

Our technical approach was based on the four slave-boson
method first introduced in Ref. \cite{kotliar}. The electron-phonon
vertex function for a finite $U$ Hubbard model
has been evaluated as linear response to an external field coupled with
charge density. In the spirit of functional integral
representation the screening of the electron-phonon function
is due to the Gaussian fluctuations of the auxiliary boson fields
around the mean-field solution. A similar study at finite $U$ was
introduced by A. Lavagna who however did not addressed the momentum
modulation of the electron-charge density response \cite{lavagna}.
A different choice for the renormalization of the $z_i$-operators was
in addition done in comparison with Refs. \cite{kotliar,lavagna} to
avoid unphysical divergences in the electron-slave boson matrix elements.
Technical details will appear in Ref. \cite{ccp}.
Based on this analytical approach,
we can now evaluate the electron-phonon function
with the same physical parameters of Ref. \cite{huang}, namely
$n = 0.88$, $\beta=2$,
where $n$ is the electron filling
and $\beta=t/k_{\rm B}T$ the inverse temperature in unit
of the nearest neighbor hopping parameter $t$.
As only slight difference with respect to Ref. \cite{huang}
we assume the incoming electron momentum ${\bf p}$ to be averaged
over the Fermi surface and the exchanged frequency to be exactly zero.
These marginal differences are expected to not significantly
affect the comparison between our results and Ref. \cite{huang}.

\begin{figure}
\centerline{\psfig{figure=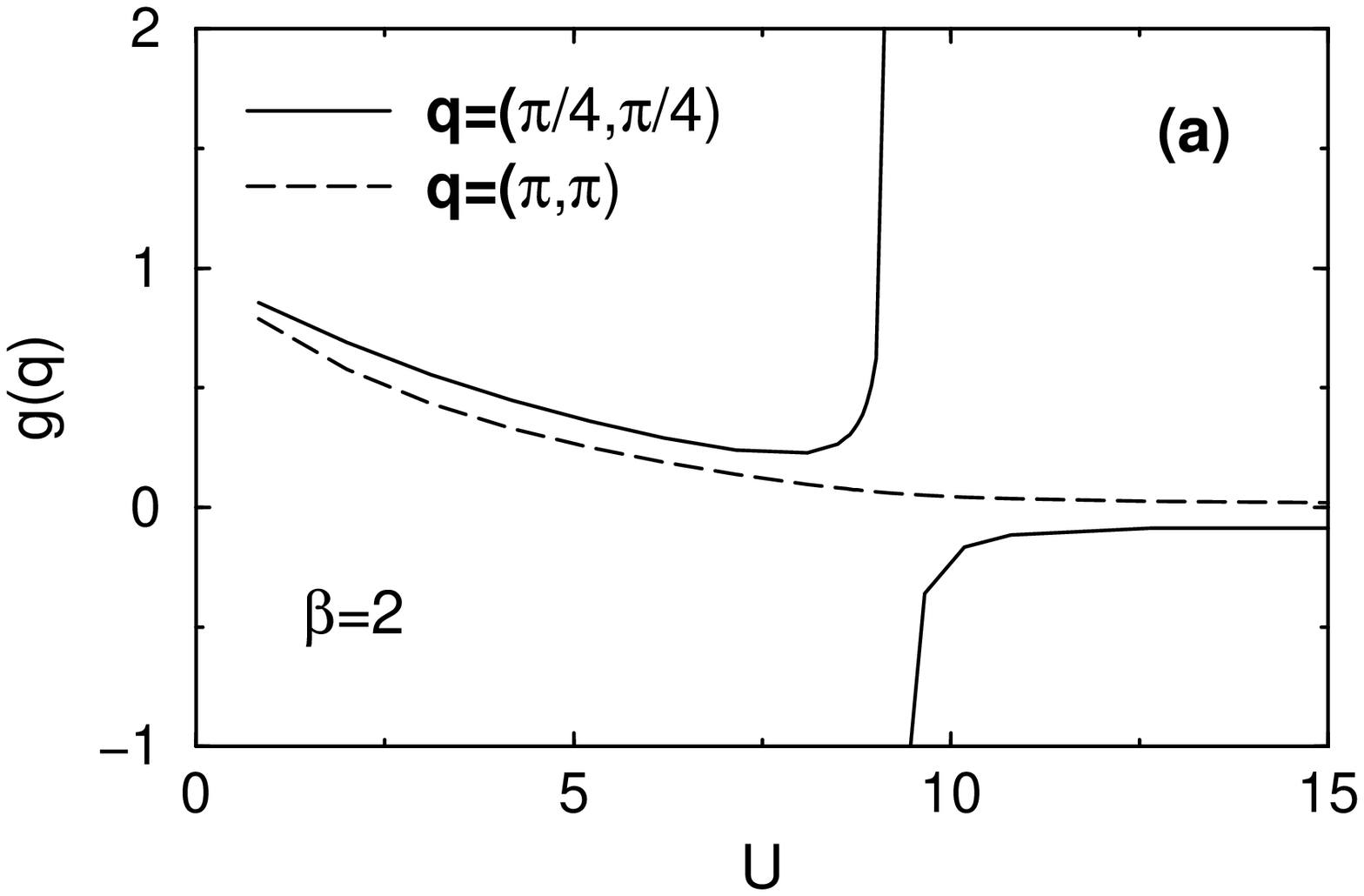,width=7.3cm}}
\centerline{\psfig{figure=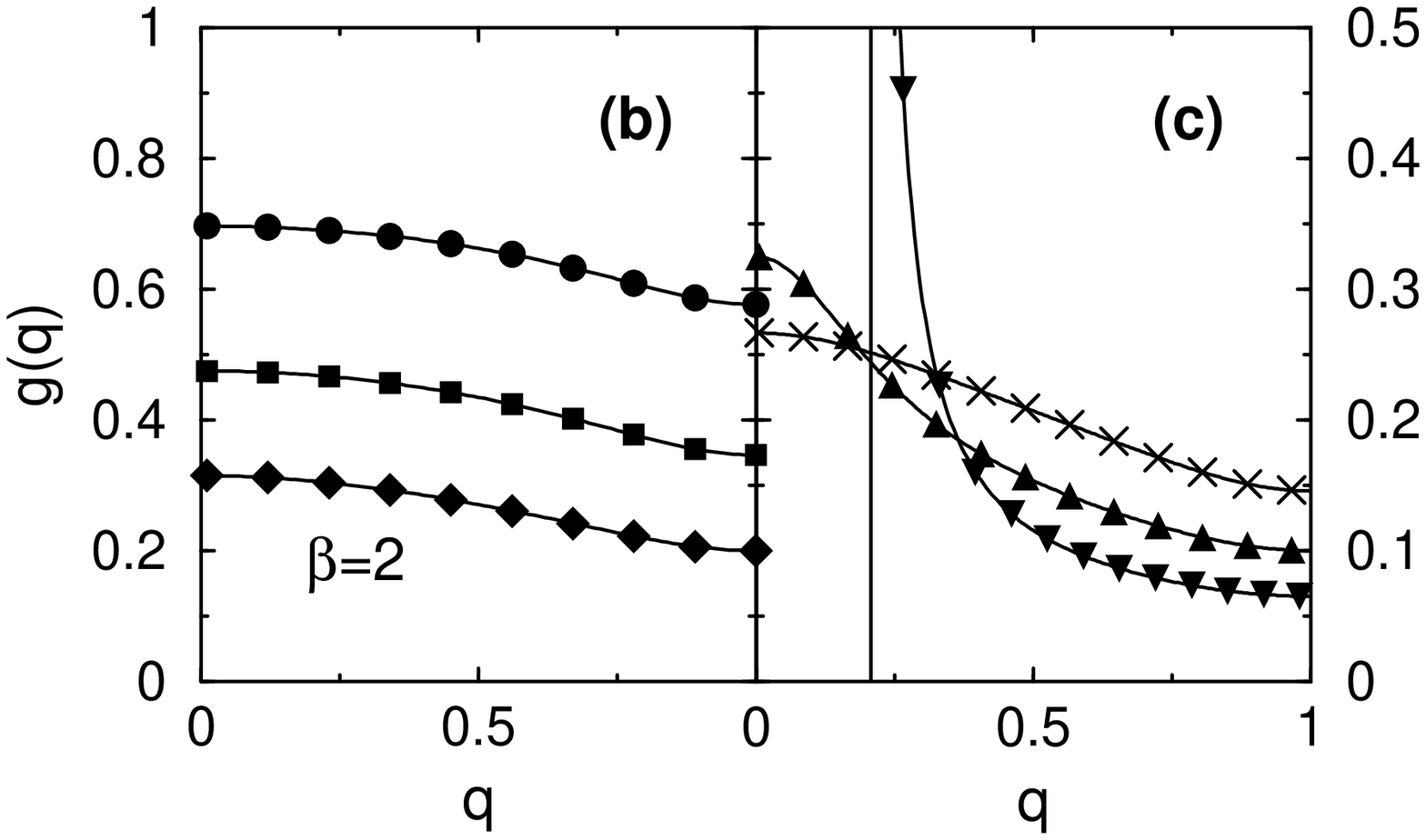,width=8cm}}
\caption{(a): Electron-phonon vertex $g({\bf q})$ as function
of the Hubbard repulsion $U$ for ${\bf q}=(\pi /4,\pi /4)$
and ${\bf q}=(\pi,\pi)$. The temperature was set here $\beta=2$
and the electronic filling $n=0.88$. (b)-(c) Plot of
$g({\bf q})$ as function of ${\bf q}=(q,q)\pi$ for $U=2,4,6$
(respectively circles, squares and diamonds in panel b) and for
$U=7,8,9$
(respectively crosses, triangles up and triangles down in panel c).}
\label{f1}
\end{figure}

In Fig. \ref{f1}a we plot the electron-phonon vertex function
$g({\bf q})$ at ${\bf q}=(\pi/4,\pi/4)$ and ${\bf q}=(\pi,\pi)$
as function of the Hubbard
repulsion $U$ for $n=0.88$ and $\beta=2$.
For ${\bf q}=(\pi/4,\pi/4)$
we note that while for relatively small $U$ the el-ph
vertex function is steady decreasing with $U$, such a behaviour
has an upturn for $U \simeq 8$ until a divergence occurs for
$U_c \simeq 9.3$. According this view one is attempted to associate
the upturn of $g({\bf q})$ as function of $U$ as an incipient transition
towards some charge instability. Note also that for
${\bf q}=(\pi,\pi)$ no charge instability is observed.

The appearance of charge instability can be also detected by looking
at $g({\bf q})$ plotted as function of ${\bf q}$ (${\bf q}=(q,q)\pi$).
The evolution of $g({\bf q})$ by varying the Hubbard repulsion $U$
is shown in Fig. \ref{f1}b,c where we see that the el-ph vertex function
is initially suppressed at small ${\bf q}$
by increasing $U$ (panel b),
then it increases as function of $U$ (panel c)
until a divergence is established.
Note that for $U=9$ a lattice instability already occurs although for a
$q$ ($\simeq 0.2$) less than $q \simeq 1/4$. This reflects the fact that
a instability for $q=0$ (phase separation) is first onset for
some critical value of $U$, and then the vector instability
is gradually shifted by further increasing of $U$.
In this perspective it is not surprising that momenta
on the Brillouin zone edge
${\bf q}=(\pi,\pi)$ are less sensitive to the increase of $U$.

The similarity between our findings and the Quantum Monte Carlo analysis
suggests that also the upturn of
$g({\bf q})$ as function of $U$ reported
in Ref. \cite{huang} could be related to the same tendency
towards phase separation or charge instabilities.
This does not imply however that phase separation is effectively established,
and it should be remarked that the actual occurrence of phase separation
in the Hubbard model is still object of debate \cite{cosentini}.
On one hand expansions
around the mean-field solution even including Gaussian fluctuations
could enforce unphysical instabilities which could disappear once
higher order fluctuations are taken into account, especially
in two dimensional systems.
On the other hand, small size cluster effects ($L \times L$)
and large temperature effects in Ref. \cite{huang}
question in principle
the generalization of the QMC results in the
thermodynamic limit ($L \rightarrow \infty$) and at low
temperatures.
Our results should thus viewed as analytical indications which
can trigger further numerical work.

As a final step, we can also employ our slave-boson analysis
to extend the range of investigation in regions of parameters not
addressed in Ref. \cite{huang}.
In particular we show that
a crucial role is played by the temperature
that is limited in QMC techniques by the sign problem and by the
requirement to be larger that the energy spectrum discretization.

\begin{figure}
\centerline{\psfig{figure=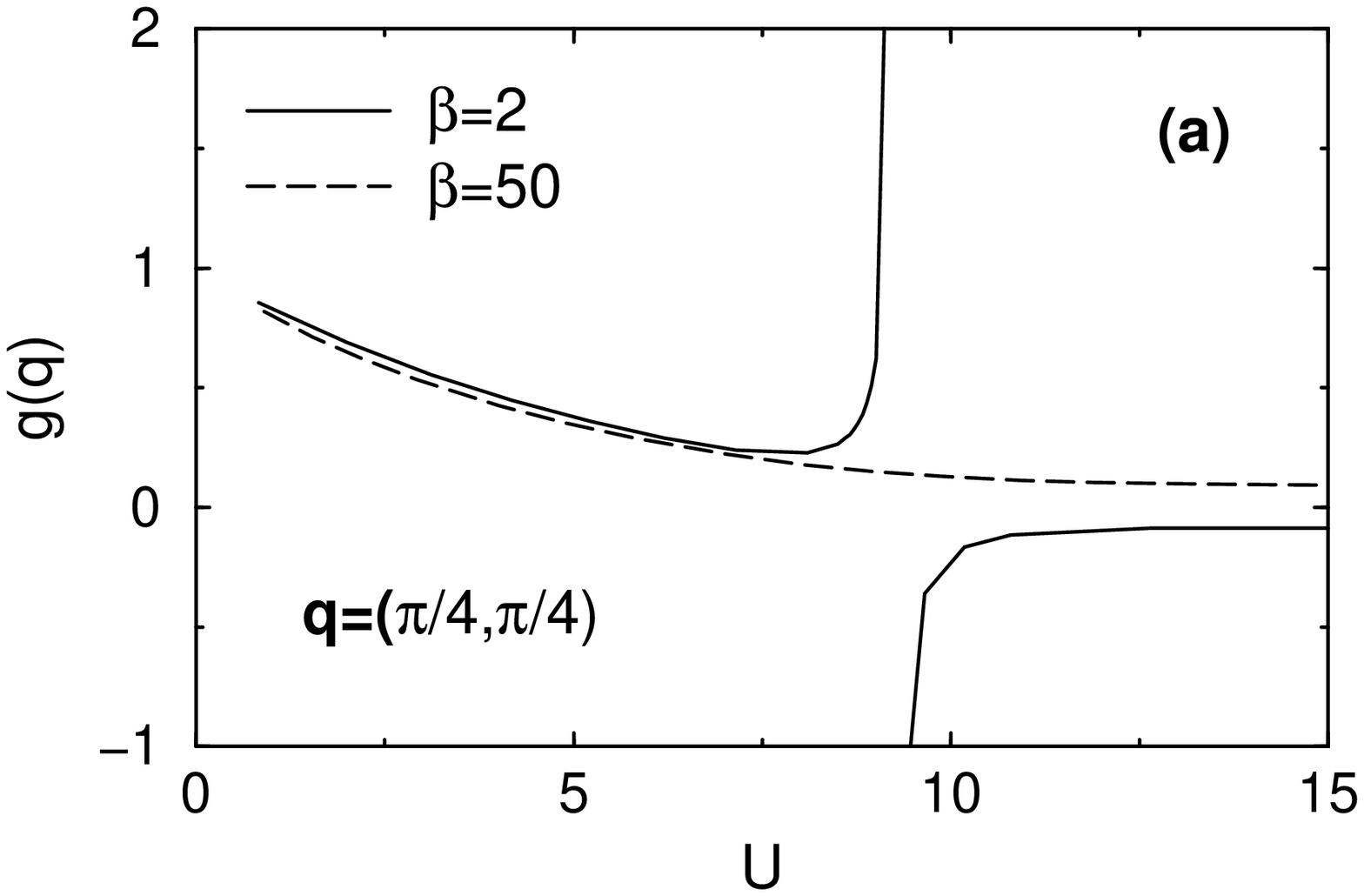,width=7.3cm}}
\centerline{\psfig{figure=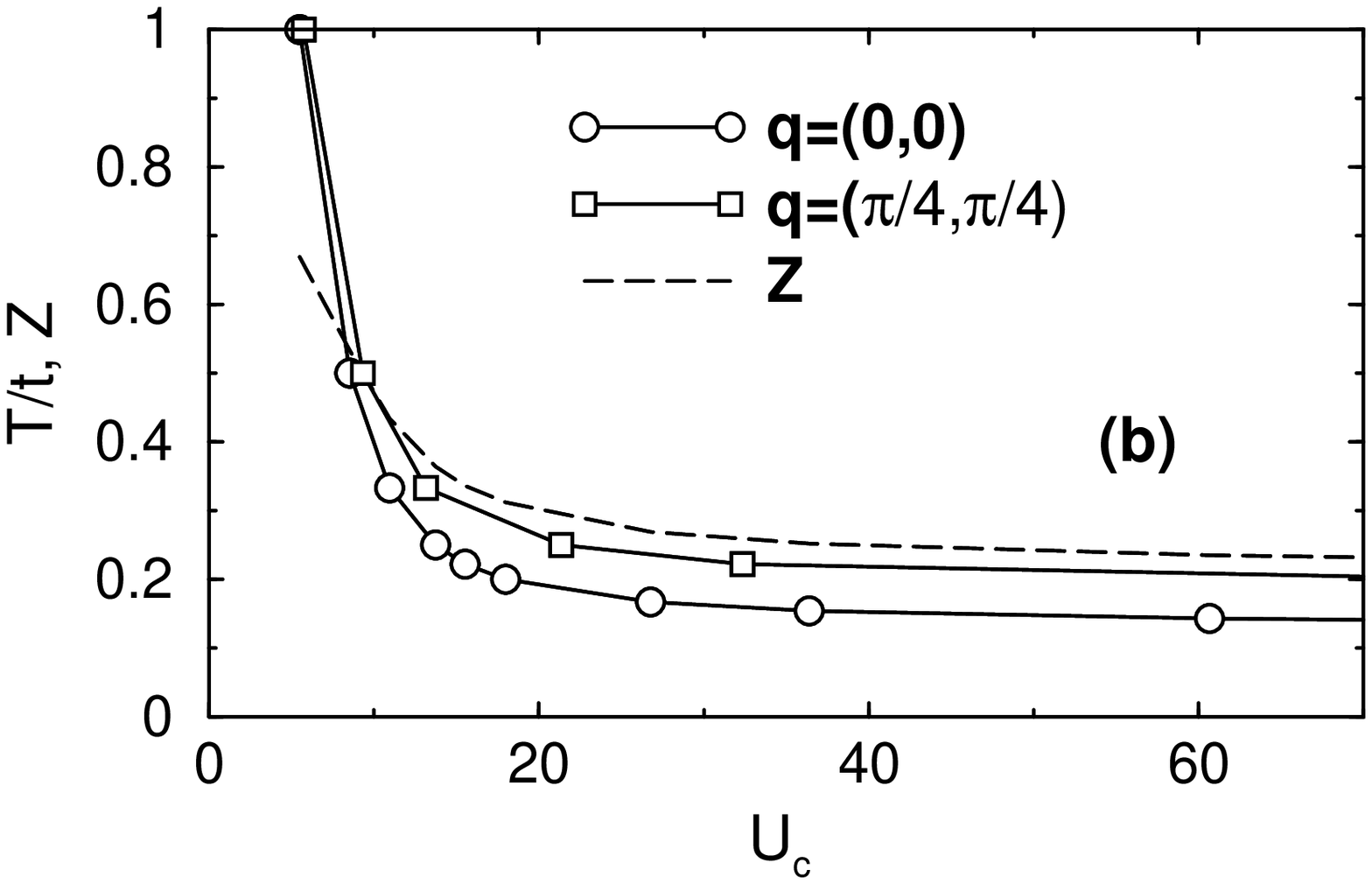,width=7.3cm}}
\caption{(a): Electron-phonon vertex $g({\bf q})$
at ${\bf q}=(\pi /4,\pi /4)$ as function
of the Hubbard repulsion $U$ for $\beta=2$
and $\beta=50$. Electronic filling $n=0.88$. (b) Phase diagram
for phase separation (${\bf q}=0$) and for
the charge ordering ${\bf q}=(\pi /4,\pi /4)$ instability
in the $T$-$U$ space. The dashed line represent the band narrowing
factor $Z$ corresponding to the phase separation instability line.}
\label{f2}
\end{figure}
In Fig. \ref{f2}a we report the dependence of the el-ph vertex function
$g({\bf q})$ at ${\bf q}=(\pi /4,\pi /4)$
as varying $U$ in the large ($\beta=2$) and small ($\beta=50$)
temperature limit. As a surprising result,
no charge instability is found in the low temperature regime
in the contrast with the high temperature range ($\beta=2$).
Same conclusions hold true for
the phase separation ${\bf q}=0$ instability (not shown in figure).
This results in thus compatible with the absence of
phase separation at zero or low temperature.

In order to understand in more detail the origin
of the phase separation instability as function
of the temperature $T$ we show in Fig. \ref{f2}b
the phase diagram in the $T$ vs. $U$ space with
respect to phase separation (${\bf q}=0$) and
to charge ordering (${\bf q}=(\pi /4,\pi /4)$).
As above mentioned,
a phase separation instability occurs
in our slave-boson calculations only above a certain temperature
$T/t \bsim 0.2$ ($\beta \lsim 5$) \cite{cz}.
As expected, finite ${\bf q}$ instability, in the absence of any long-range
Coulomb repulsion, is prevented by the occurrence of phase separation
at ${\bf q}=0$ in the whole phase space.
A interesting insight comes from the comparison of the critical
temperature $T_c$ at which the instability towards phase
separation occurs with the band narrowing factor $Z$
due to the correlation effects \cite{kotliar,lavagna,ccp}
(dashed line
in Fig. \ref{f2}b).
The similar dependence on $U$ of $T_c$ and $Z$ points out
that the onset of phase separation by increasing temperature
is ruled by the comparison between the temperature $T$ and
the effective bandwidth $W=Z 8t$ energy scales.
In particular phase separation is established when $T$ become
is of the same order of with $Zt$ ($=W/8$).
Once again we stress that the
phase separation instability found by
our slave-boson calculations which include Gaussian
fluctuations could be washed out
when higher order fluctuations are taken into account,
so that it should be regarded as indicative of tendency towards
this kind of instability.
Numerical work based on Quantum Monte Carlo techniques
will help to answer about the effective role of phase separation
or charge ordering in the Hubbard model.

%\end{acknowledgments}

\vskip\baselineskip

\noindent{B. Cerruti$^a$, E. Cappelluti$^{b,*}$ and L. Pietronero$^a$}

\baselineskip = 10 pt
\indent
{\small $^a$Department of Physics}
\newline\indent
{\small University of Rome ``La Sapienza''}
\newline\indent
{\small P.le Aldo Moro 2, 00185 Rome, Italy}
\vskip\baselineskip

{\small $^b$Centro Ricerche ``Enrico Fermi'' of Rome}
\newline\indent
{\small v. Panisperna 89a, c/o Compendio Viminale}
\newline\indent
{\small 00184 Rome, Italy}
%\newline\indent
%{\small {\tt emmcapp@pil.phys.uniroma1.it}}
\vskip\baselineskip

\noindent
{$^*$Corresponding author: 
{\tt emmcapp@pil.phys.uniroma1.it}}


\end{document}